\documentstyle{aipproc}

\begin{document}

\title{Optical/Ultraviolet Continuum Emission Theory in Radio Quiet
Quasars and Active Galactic Nuclei}
 
\author{Omer Blaes}
\address{Physics Department, University of California, Santa Barbara, CA 93106}

\maketitle

\begin{abstract}
Accretion disk models still do not provide a satisfactory explanation of
the optical/ultraviolet continuum observed in Seyferts and quasars.  Substantial
theoretical progress has been made in understanding one aspect of the problem:
the dearth of spectral features at the Lyman limit.  Promising solutions have
also been proposed to explain the surprising observations of large
polarization in the Lyman continuum observed in some sources.  I review
the recent progress in this field, and try to point out future research
directions which would be fruitful in trying to obtain a complete,
self-consistent model of the continuum emitting regions.
\end{abstract}

\section*{Introduction}

Thermal emission from accretion disks has long been the standard paradigm
for the optical/ultraviolet (OUV) continuum in quasars and active galactic
nuclei (AGN).  However, as noted in the talk by Koratkar in this conference
\cite{kor97a}, the simplest, bare, geometrically thin, optically thick
accretion disk models face a number of severe problems when confronted with
observations.  These include the observed absence of absorption or emission
edges at the Lyman limit in most sources \cite{ant89,kor92}, the continuum
polarization \cite{ant92,ant96}, the
continuum spectral energy distribution \cite{lao90}, phased
optical/ultraviolet variability \cite{all85,cut85,cou91},
and microlensing constraints in the Einstein Cross \cite{rau91,cze94}.
In addition, there are
issues around connecting the ultraviolet continuum to photoionization models
of the emission line regions \cite{kor97}, and how the OUV continuum is related to
the soft and hard X-rays observed from these sources.  As Koratkar emphasizes
in her talk, a theorist's job is to design a predictive model which explains
all the observed phenomenology simultaneously, not just in a piecemeal
fashion.  Such a model does not yet exist, although progress is definitely
being made.  In this paper I will review some of the progress, concentrating
in particular on the Lyman edge problem because it sheds light on many of
the other issues.

\section*{The Lyman Edge Problem}

The Lyman edge problem arose from early calculations of AGN accretion disk
spectra by Kolykhalov \& Sunyaev (1984)\cite{kol84}.  These authors adopted
the reasonable approach of taking existing stellar atmosphere libraries at
that time \cite{kur79,tsu76,tsu78}, assigning a spectrum to each radius
of an alpha disk \cite{sha73} by using an atmosphere model which had the same 
effective temperature and surface gravity, and then integrating over the
disk to produce the resulting spectrum.  All the disk models they explored
had very large absorption edges at the Lyman limit, and they noted
that these disagreed with observations of bright quasars at the time.
This discrepancy motivated much of the later theoretical and observational
work on the Lyman edge region of AGN spectra, but it is important to note
the limitations of these early calculations.  The stellar atmosphere
libraries which were used completely neglected non-LTE effects.  These
can be very important in the Lyman edge region because the large
hydrogen photoionization opacity generally means the Lyman continuum
originates high in the atmosphere where densities are low and radiative
transition rates can dominate collisional transition rates.  Even more
important, the atmospheres which exist in those libraries generally have
high densities compared to what is possible in high luminosity accretion
disks.  This limited the range of parameter space which Kolykhalov \&
Sunyaev could explore to low alpha parameter ($\alpha<10^{-2}$), high
photospheric density models.  Because the neutral fraction of hydrogen
becomes larger at high densities, this resulted in very large Lyman edge
absorption opacity, thereby producing very large absorption edges in these
disk models.

As discussed in more detail by Koratkar\cite{kor97a}, spectral features
near the Lyman limit (whether in emission or absorption) are very rare in
quasars, except in cases where there is clearly evidence for intervening
absorbers \cite{ant89,kor92}.  There {\it are} cases of weak features which
are consistent with being intrinsic to the quasar, but only in $\sim10$\% of
quasars studied \cite{kor92}.  Intervening absorption by the intergalactic
medium is less of a problem for low redshift Seyferts, but one must then
contend with the absorption due to our own Galaxy.  Hopkins Ultraviolet
Telescope data generally show no features at the intrinsic Lyman limit
of low redshift type 1 Seyferts \cite{kri97}, including
NGC~5548.  This is particularly disturbing as NGC~5548 has a broad iron
K$\alpha$ line, which is strong evidence for a relativistic accretion disk
\cite{mus95,nan97}.  There is some absorption at slightly shorter wavelengths,
but this could conceivably be due to high $n$ Lyman series lines from the
Galaxy.  Zheng et al. (1995)\cite{zhe95}
find that an intrinsic Lyman edge provides a good fit to the data for
the type 1 Seyfert Mrk 335.  More data and careful modeling is clearly needed.

The Hubble Space Telescope (HST) composite quasar spectrum of Zheng et al.
(1997)\cite{zhe97} shows
at most only a very shallow absorption feature at the Lyman limit, but it
{\it does} show a spectral break near there.  The spectrum shortward of the
Lyman limit can be fit with a power law, and extension of that power law lines
up approximately with the ROSAT composite spectrum of Laor et al.
(1997)\cite{lao97}.  This spectral break near the Lyman limit actually causes
problems in another area of AGN phenomenology, that of the emission line
regions. If the line emitting gas sees the same photoionizing continuum as
we observe in the composite, then there are too few photons to produce the
observed He~II emission lines \cite{kor97}.  It is important to remember that
Zheng et al. had to correct for
the Lyman $\alpha$ forest and Lyman continuum absorption by intervening
material.  In addition, they did not correct for internal extinction, which
could be very important in the ultraviolet.  (This may nevertheless be okay
given the lack of any obvious 2200\AA~ feature.)  The composite spectrum may
indicate that many quasars do have absorption at the Lyman limit combined
with e.g. Comptonization, as the authors suggest.  It would be nice to see
this spectral energy distribution in an individual source, however.

There have been essentially three theoretical approaches to solving the Lyman
edge problem within the standard optically thick, geometrically thin accretion
disk paradigm: more sophisticated stellar atmosphere modeling, Comptonization
by a hot corona, and relativistic smearing.

\subsection*{Stellar Atmosphere Effects}

Low densities in the disk photosphere will reduce the ratio of Lyman continuum
absorption opacity to scattering opacity, and thereby decrease the flux
difference at the absorption edge, or even drive it into emission \cite{cze90}.
Ab initio radiative transfer calculations confirm this behavior
\cite{col93,shi94,hub97a,sin97b}, showing that as the maximum effective
temperature
and/or the ratio of luminosity to Eddington luminosity increase, Lyman
absorption edges become weaker, then disappear, become moderate emission
edges, and finally become weak emission edges as the hydrogen becomes fully
ionized and recombination rates are low.  Non-LTE effects,
where the ionization states of hydrogen are determined
at least partially by the radiation field rather than just by collisions,
produce similar results \cite{sun89,shi94,sto94,hub97a}.  (They can also
greatly enhance the He~II Lyman continuum \cite{hub97a}, which may help
photoionization models of the emission line regions \cite{kor97}.)  To
completely get rid
of the edge in a given atmosphere using these effects would require
fine-tuning, but it is clear that the overall edge produced from an integration
of diverse atmospheres at different disk radii, some with absorption edges
and some with emission edges, might generically produce a small net effect.

Most of the radiative transfer calculations in AGN accretion disk theory
have been fairly crude or incomplete, and it is only recently that
modern, state of the art, stellar atmosphere codes have been applied to the
problem.  A crucial effect that has generally been neglected in the past
because of its complexity is metal line blanketing.  Hubeny \& Hubeny have
shown at this conference \cite{hub97b} that metal lines can produce much
stronger spectral
features in an individual atmosphere in the Lyman edge region than the Lyman
edge itself.  In fact, judging from their plots, an observer
would be hard-pressed to see any Lyman edge feature, even if they could
observe the atmosphere directly without the effects of Comptonization and
relativistic smearing discussed below.

All these stellar atmosphere models neglect the effects of external illumination
of the atmosphere.  AGN and quasars emit X-rays, and in Seyferts, X-rays
carry a substantial fraction of the bolometric luminosity.  Indeed, in the
most extreme form of the currently popular two phase disk/corona accretion
model \cite{haa91}, {\it all} the accretion power is initially converted
into heat in the corona above the disk.  Disk heating would therefore take
place externally, rather than internally.  Real sources probably have a
combination of internal and external heating, such as in the patchy corona
model of Haardt, Maraschi, \& Ghisellini (1994)\cite{haa94}.  Sincell \&
Krolik (1997a)\cite{sin97a} have done the best job so far of calculating the
resulting spectrum from an X-ray illuminated disk, starting with the
extreme assumption of total external heating.  The lack of
internal, viscous heating results in a much lower disk scale height and
therefore a much larger photospheric density.  As a result, they found very
substantial Lyman edges in emission, in disagreement with observation.
This calculation would be worth doing even better, however.  The X-rays were
incorporated just through a depth-dependent heating function, and were therefore
not included as part and parcel of the radiation field in the radiative
transfer and atmosphere structure calculation.  In addition, the calculation
was done in LTE, and therefore X-ray and UV photoionization effects on
the ground state population of hydrogen were neglected.  Even so, the trend
of large edges produced with high photosphere densities may be robust, and
may spell trouble for the simplest two-phase models.

It is perhaps useful to introduce a reality check here by comparing how
well stellar atmosphere models do in reproducing the Lyman edges in real
stars, which we know are optically thick!  The B2~II star $\epsilon$~CMa
exhibits a substantial Lyman absorption edge, but much less than predicted
by both LTE and non-LTE models with line blanketing\cite{cas95}.  In fact,
the Lyman continuum in this star exceeds that predicted by the stellar
atmosphere models by a factor $\sim30$!  It is interesting that one proposed
solution of this problem involves external illumination of the star's
atmosphere by X-rays \cite{hub95}.  Stellar atmosphere models also fail
to reproduce the spectrum of the B1~II-III star $\beta$~CMa in all wavelength
regions \cite{cas96}.

\subsection*{Comptonization}

If X-rays are produced by Comptonization in a hot corona above the disk,
then the optical/ultraviolet spectrum will itself be modified in the
process.  Comptonization can substantially reduce a Lyman absorption edge
by scattering low energy photons above the edge, thereby filling it up
\cite{cze91,lee92,hsu97}.  Downscattering by a hot corona can also reduce the
magnitude of an emission edge discontinuity, although not as dramatically for
the same coronal optical depth and temperature because of the tendency for
upscattering in a hot corona \cite{hsu97}.  In either case, a spectral
break in the continuum slope is generically produced for modest optical
depths, and this has been used to fit the HST composite spectrum \cite{zhe97}.

Antonucci (1992)\cite{ant92} has criticized this solution to the Lyman
edge problem on the grounds that the modest scattering depths required would
impart substantial polarization everywhere in the optical/ultraviolet, which
is not observed.  In fact Compton scattering generally produces less
polarization in a featureless continuum source than Thomson scattering, but
nevertheless substantial polarization does result if an edge is successfully
smeared out \cite{hsu97}.  It is likely that a Comptonizing corona is
powered through magnetic fields, and if these are anywhere near equipartition
strength with the radiation field, then Faraday rotation can greatly
reduce the polarization (see below).

\subsection*{Relativistic Doppler Shifts and Gravitational Redshifts}

Many authors have invoked the highly relativistic orbital velocities and
varying gravitational redshift with distance from the hole to smear out
the Lyman edge (e.g.\cite{sun89,lao89,lao92,lee92,col93,shi94,weh97}).
Not surprisingly, Kerr holes are more effective at smearing edges than
Schwarzschild holes because of the deeper potential wells.  In addition,
edge-on disks are more effective than face-on disks.  However, it should be
born in mind that type 1 objects are probably viewed more face-on than
edge-on, at least according to the unified model \cite{ant93} and the
inferences from the iron K$\alpha$ lines \cite{nan97}.  It is interesting
to note that the paper by
Kolykhalov \& Sunyaev (1984)\cite{kol84}, which first articulated the Lyman
edge problem as discussed above, did in fact include relativistic smearing
by a Schwarzschild hole.  This failed to remove
the huge Lyman discontinuities which were produced by their dense atmosphere
models.

Agol, Hubeny, \& Blaes have presented results at this conference \cite{ago97b}
of line-blanketed stellar atmosphere models (cf.\cite{hub97b}) folded through
the relativistic transfer function of Agol (1997)\cite{ago97a}.  The individual
atmosphere models are not yet fully self-consistent, and more parameter space
needs to be explored, but the results so far indicate very weak Lyman edges.

\section*{Polarization}

A long-standing criticism of geometrically thin, optically thick accretion
disk models is that the emerging radiation should be polarized.  Chandrasekhar
(1960)\cite{cha60} showed that an optically thick, pure electron scattering
atmosphere produces a polarization of up to 11.7\% parallel to the atmosphere
plane.  Real accretion disks, however, also have significant
absorption opacity \cite{lao90b} and may not be exactly planar \cite{col90}.
In addition, if photospheric magnetic fields are anywhere near equipartition
strength, then the optical radiation field could be completely depolarized
by Faraday rotation \cite{gne78,bla90,mat93,beg94,ago96,lee97,ago97c}.  The
observed optical polarization may be due to scattering by material further
out from the central engine (e.g.\cite{kar95}).

The Lyman limit region has again turned out to be extremely interesting with
regard to polarization.  As reviewed in more detail by Koratkar in this
conference \cite{kor97a}, some quasars show steep rises in polarization
shortward of the Lyman limit \cite{imp95,kor95}, in one case to $\sim20${\%}!
The samples are not statistically homogeneous, but there appears to
be an association between the presence of a partial Lyman absorption edge
and significant polarization shortward of the edge, although this is not
always the case \cite{kor97b}.

Substantial anisotropy in the radiation field (limb darkening) at short
wavelengths can be produced in a cool disk atmosphere because of the steep
increase in the source function as the temperature rises with depth.  This
effect can produce a sharp rise in polarization shortward of the Lyman limit
which qualitatively agrees with the observations \cite{bla96}, although
quantitative agreement with the data has not yet been obtained.  Relativistic
smearing of the Lyman edge also blueshifts and reduces the rise in
polarization \cite{ago97a,shi97}.  It is possible to smear out the edge in total
flux and maintain a steep polarization rise if the optically thick portion
of the disk does not extend all the way down to the innermost stable circular
orbit, e.g. if there is an advection dominated accretion flow in the inner
region.  Comptonization also tends to smear out a steeply rising polarization
emerging from an underlying disk photosphere, although again, there are
regions of parameter space where the edge is substantially reduced in total
flux while maintaining a steep polarization rise \cite{hsu97}.

Shields, Wobus, \& Husfeld (1997)\cite{shi97} have recently come up with an
interesting model which, while {\it ad hoc}, produces excellent quantitative
agreement with the observations.  They assume that each annulus of the disk
locally produces a spectrum with a large Lyman edge in absorption.  In
addition, they assume that the local polarization longward of the edge
is near zero, while shortward of the edge it is {\it arbitrarily} large.
After folding through the relativistic transfer function, this model can
fit the data with few free parameters.  Shields et al. suggest that
this toy model might be produced by an optically thin, ionized region of
the disk which emits Lyman continuum photons which are then polarized by
scattering.  This ionized gas might be produced by photoionization by X-rays
from a corona, or perhaps be an optically thin, inner region of an otherwise
optically thick disk.  We have also found recently that Comptonization
of an absorption edge reproduces the features of their toy model quite
naturally \cite{hsu97}.  The polarization shortward of the edge is negative,
in the sense that the plane of polarization is parallel to the disk axis, in
contrast to the polarization produced by limb darkening invoked by
\cite{bla96}.  However, it remains to be seen whether a spectrum produced by
a self-consistent atmosphere/corona folded through the relativistic transfer
function can truly reproduce the data.

\section*{Alternatives to the Optically Thick Accretion Disk Paradigm}

The problems with the standard accretion disk model have led some authors
to explore alternatives based on optically thin thermal emission
(e.g.\cite{fer88,fer90,cel92,bar93,col96,kun97}).  The fact that single
temperature thermal bremsstrahlung radiation provides a simple quantitative
fit to the observed continuum spectral energy distributions should perhaps
not be dismissed lightly.  However, optically thin models also generally
produce very significant bound-free and bound-bound emission, and therefore
suffer similar problems.  Again, Comptonization and relativistic smearing
can be invoked to get rid of these spectral features.  The broad iron
K$\alpha$ lines \cite{nan97} and X-ray reflection humps \cite{nan94} observed
in Seyferts are
a strong argument for the presence of slabs of optically thick material,
such as would be present in the standard accretion disk paradigm.
Hybrid models of magnetically confined clouds above an accretion disk have
also been considered (e.g.\cite{kun97}).

It may very well be that a multiphase medium will be necessary to explain
the entire spectral energy distribution from optical to hard X-rays of
Seyferts and quasars.  Preliminary work by Magdziarz et al. (1997) \cite{mag97}
invokes such a mixture of a cold disk, hot X-ray emitting
plasma, and an intermediate temperature, optically thick, Comptonizing phase
to explain the OUV, soft X-ray excess, and
hard X-rays in NGC~5548.  More simultaneous, multiwavelength variability
campaigns are necessary to provide us with the data required to resolve the
geometry of the continuum emission regions and to sort out the energetics
of the reprocessing.

\section*{Conclusions}

In the particular problem of the absence of Lyman edge features in quasars,
theory and observation appear to be converging, and I believe that realistic
accretion disk models will have little trouble in explaining the absence
of such edges in total flux in the majority of objects.  Whether this will
extend to Seyfert galaxies remains to be seen, and requires additional
spectroscopy data in the Lyman limit region of sources at moderate redshifts
(hopefully which also exhibit broad iron K$\alpha$ lines)
as well as more theoretical calculations of the effects of external illumination
by X-rays.  There are promising solutions proposed for the polarization
observations around the Lyman edge, although a complete model based on
first principles has yet to be constructed.

While all this is good news, it must be remembered that the Lyman edge problem
is only one aspect of the overall theoretical problem.
For example, Sincell \& Krolik (1997b)\cite{sin97b} have correctly noted that
in order to have a standard, viscously heated, optically thick disk be hot
enough to drive the local Lyman edge into emission in the innermost radii,
the near ultraviolet spectral energy distribution would be close to the long
wavelength $F_\nu\propto\nu^{1/3}$ prediction of a standard blackbody disk.
This is much harder than observed AGN continuum spectra, assuming one accepts
(as I do) that the
infrared emission is due to dust and therefore cannot extend underneath
the big blue bump.  There is still no accretion disk model which satisfactorily
fits all the energetically important regions of an AGN spectrum simultaneously.

Variability campaigns in Seyfert galaxies have demonstrated that the OUV and
X-ray continua are directly connected through reprocessing.  The recent
detections of delays between optical and ultraviolet bands in the recent
monitoring campaign of NGC~7469 \cite{wan97} are tremendously exciting, and
together with X-rays, may shed light on this problem.  Well-sampled,
simultaneous, observing campaigns which cover all the energetically important
regions of the AGN spectrum (including the ``soft X-ray excess'') are likely
to produce significant progress in our understanding of the geometry of the
continuum emission regions in the near future.

\section*{ACKNOWLEDGMENTS}

I have benefited enormously from collaborations and/or conversations with
Eric Agol, Robert Antonucci, Robert Goodrich, Ivan Hubeny, Veronika Hubeny,
Chia-Ming Hsu, Anuradha Koratkar, Julian Krolik, Pawe\l~ Magdziarz, Greg
Shields, and Mark Sincell.  This work was supported by NSF grant AST~95-29230.

\end{document}